\documentclass[twocolumn,eqsecnum,aps,epsfig]{revtex4}
\usepackage{epsfig}

\def\rhovec{\mbox{\boldmath $\rho$}}

\begin{document}

\title {Five-body calculation of resonance and scattering states
of pentaquark system 
}

\author{E.\ Hiyama$^a$, M. Kamimura$^b$, A. Hosaka$^c$, H.Toki$^c$
 and M. Yahiro$^b$}

\address{$^a$Department of Physics,
Nara Women's University, Nara 630-8506, Japan}

\address{$^b$Department of Physics, Kyushu University,
Fukuoka 812-8581, Japan}

\address{$^c$Research Center for Nuclear Physics (RCNP),
Osaka University, Ibaraki, 567-0047, Japan}

%


\begin{abstract}
Scattering problem of the $uudd{\bar s}$ system, 
in the standard non-relativistic quark model of Isgur-Karl, 
is solved for the first time,  by treating 
the large five-body model space 
including the $NK$ scattering channel 
accurately with the Gaussian expansion method 
and the Kohn-type coupled-channel variational method.  
The two-body interaction that reproduces observed properties 
of ordinary baryons and mesons is applied to the pentaquark system
with no additional adjustable parameter.
The five-body wave function calculated has the correct asymptotic form 
in its the scattering-channel component and describes 
$qq$ and $q{\bar q}$ correlations  properly. 
The $NK$ scattering phase shift calculated shows 
no resonance in the energy region of the reported pentaquark 
$\Theta^+(1540)$, that is, at $0-500$ MeV above the
$NK$ threshold ($1.4-1.9$ GeV in mass). 
The phase shift does show two resonances 
just above $500$ MeV: 
a broad $\frac{1}{2}^+$ resonance with a width of 
$ \Gamma \sim\!\!~110$ MeV located 
at $\sim\!\!~520$ MeV ($\sim\!\!~2.0$ GeV in mass) and 
a sharp $\frac{1}{2}^-$ resonance with 
$\Gamma=$0.12 MeV at 540 MeV.  
Properties of these states are discussed. 
\end{abstract}

\maketitle

\section{Introduction}

The observation of a signal of a narrow resonance 
at $\sim\!\!~1540$ MeV with 
$S = +1$ by the LEPS group \cite{Nakano03,Dzierba05}, 
now called  $\Theta^+(1540)$,  
triggered a lot of theoretical works 
on multi-quark systems \cite{Oka04}, 
although further experimental reexamination is still needed
to establish the state. 
The question whether the multi-quark baryon $\Theta^+(1540)$ 
really exists or not 
is then one of current issues in hadron physics. 
In order to answer the question theoretically, one has to
nonperturbatively  evaluate the mass  and 
the decay width of $\Theta^+(1540)$, 
namely of the $uudd{\bar s}$ resonance. 
All nonperturbative analyses made so far, however, did not impose 
any proper boundary condition 
to the $NK$ scattering component of the pentaquark state. 
At the present stage, the non-relativistic quark model provides 
 a nonperturbative framework 
which makes it possible to 
impose a proper boundary condition to the $NK$ scattering component. 
 
In this paper, we take the standard quark model 
of Isgur-Karl~\cite{Isgur71,Isgur83}. 
The Hamiltonian consists of the confining potential of 
harmonic oscillator type and the color-magnetic 
spin-spin interaction.  
As shown later, the Hamiltonian satisfactorily 
reproduces 
observed properties of ordinary baryons and mesons. 
The same Hamiltonian is applied to the pentaquark
with no adjustable parameter. 

Resonant and non-resonant states of 
the $uudd{\bar s}$ system are nonperturbatively derived 
with the Kohn-type coupled-channel variational method \cite{Kami77} 
in which a proper boundary condition is imposed to the $NK$ scattering 
channel and the total antisymmetrization between quarks is explicitly taken 
into account. 
Reliability of the method for 
scattering problem between composite particles 
was shown by one of the authors (M.K.)~\cite{Kami77}, and 
actually it was already applied to $qqq-qqq$ scattering~\cite{Oka81}.

The coupled-channel variational method is accurate, 
only when the five-body dynamics in the interaction 
region is solved precisely. As such a method, 
we use the Gaussian expansion method (GEM)~\cite{Hiyama03}. 
GEM is one of the most reliable few-body methods proposed 
by two of the present authors (E.H. and M.K.) and their collaborators. 
The method was successfully applied 
to various types of three- and
four-body systems \cite{Hiyama03}.  
For instance, the mass of antiproton~\cite{particle2004} 
was evaluated precisely, i.e. with eight-digit accuracy, 
by comparing the three-body GEM calculation~\cite{Kino04} 
with CERN's high-resolution laser 
spectroscopy data \cite{Hori03} on highly excited 
three-body resonance states of antiprotonic 
helium atom ($^4{\rm He}^{++} +  {\bar p} + e^- $).

\begin{figure*}[htb]
\begin{center}
\epsfig{file=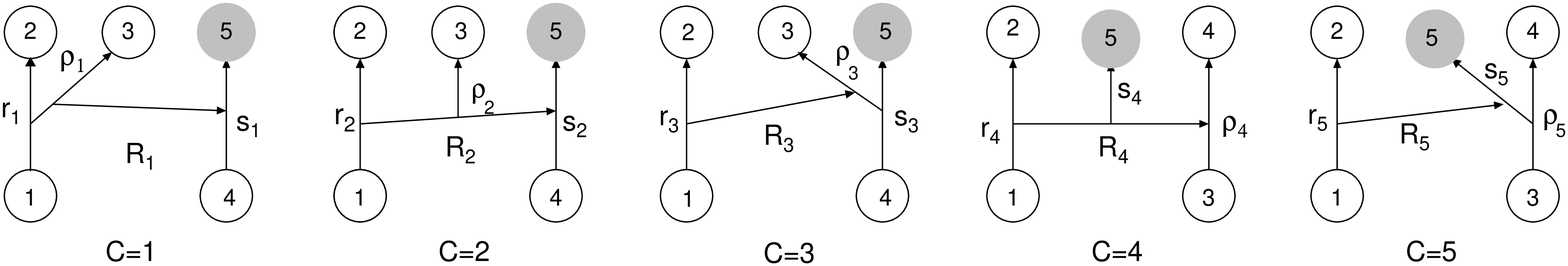,scale=0.35}
\end{center}
\caption{Five sets of Jacobi coordinates among five quarks. 
Four $u, d$ quarks, labeled by particle $1-4$, are to be 
antisymmetrized,
while particle 5 stands for ${\bar s}$ quark. 
Sets $c=4, 5$ contain  two $qq$ correlations,  
while sets $c = 1-3$ do both $qq$ and $q{\bar q}$ correlations. 
Sets $c=4, 5$ describe molecular configurations and 
sets $c=1-3$ does connected ones, as shown in the text. 
The $NK$ scattering channel is treated with $c=1$.
} 
\label{fig:penjacobi} 
\end{figure*} 

The five-body model space considered here consists of 
two parts: the asymptotic-region part describing $NK$ scattering 
and the interaction-region part. 
In this paper, the interaction-region part is accurately described 
as a superposition of an enormous number of $L^2$-type basis functions. 
As internal coordinates of the five-body system, we take 
five types of Jacobi-coordinate sets shown in Fig.~\ref{fig:penjacobi}. 
Advantages in using several
rearrangement Jacobi-coordinate sets  
simultaneously are 
reviewed in \cite{Hiyama03}. 
This setting can accommodate a wide (practically sufficient) 
model space, as shown later. Sets $c=4$ and 5 of Fig.~\ref{fig:penjacobi}
contain internal coordinates of two $qq$ pairs, so 
these sets can treat $qq$ correlations properly with the basis functions 
of the internal coordinates. 
Similarly, sets $c=1-3$ are 
convenient for handling $qq$ and $q{\bar q}$ correlations. 
In general, the two-body correlation in the color-singlet
$(q{\bar q})_{\bf 1}$ pair is twice as strong as that 
in the color-antitriplet $(qq)_{\bf {\bar 3}}$ pair 
due to the $SU(3)$ color operator. 
Sets $1-3$ contain two $(q{\bar q})_{\bf 1}$ pairs, while 
sets $4$ and 5 do a $(q{\bar q})_{\bf 1}$ pair and a 
$(qq)_{\bf {\bar 3}}$ pair. Thus, 
use of sets $1-3$ is indispensable. 
In principle, 
four more Jacobi-coordinate sets are possible, 
but these are much less important 
since they contain only one $qq$ or $q{\bar q}$ pair. 

The five Jacobi-coordinate sets
can be classified into $c=1-3$ and $c=4, 5$.
We call the latter two sets ($c=4, 5$) the "connected" configurations 
in the sense that they contain no color-singlet cluster;  
while we call the former sets ($c=1-3$) the "molecular" ones 
as they are composed of color-singlet clusters. 
The explicit definitions of these configurations are shown later. 
Obviously, the $NK$ scattering component, described
with $c=1$, is molecular since 
it contains color-singlet $qqq$ and $q{\bar q}$ clusters. 

We consider three spin-parity states, 
$J^\pi =\frac{1}{2}^-$ and $J^\pi=\frac{1}{2}^+$ and $\frac{3}{2}^+$, 
with a common isospin $T=0$. 
In this model Hamiltonian,
the $J^\pi=\frac{1}{2}^+$ and $\frac{3}{2}^+$ states are degenerate
in the absence of spin-orbit forces between quarks.
(Such a spin-orbit interaction leads to mass splitting
between $J^\pi =\frac{1}{2}^+$ and $\frac{3}{2}^+$
and influence on their phase shift {Ref.
\cite{hashimoto,hyslop}.).
Thus, our analysis is focused on the 
$J^\pi=\frac{1}{2}^+$ state. 
As shown later, the calculated phase shift 
exhibits no resonance when only the $NK$ channel
is taken and then no excitation of $N$ and $K$ is 
taken into account.
Therefore, of importance is 
whether the five-body dynamics in the interaction region 
can generate any resonance in the intermediate stage of
scattering, particularly in the
energy region of $\Theta^+(1540)$. 
Furthermore, if a resonance appears, 
of interest is whether it has the connected configuration 
or the molecular one.

\section{Model Hamiltonian}

Hamiltonian of the standard non-relativistic quark model 
of Isgur-Karl~\cite{Isgur71,Isgur83} is
\begin{equation}
 H =\sum_i \left( m_i + {{{\bf p}_i^2} \over {2m_i}} \right)
- T_G + V_{\rm Conf} + V_{\rm CM}  \;,
\label{hamiltonian}
\end{equation}
where $m_i$ and ${\bf p}_i$ are the mass and momentum 
of $i$th quark and 
$T_G$ is the kinetic energy of the center-of-mass system.  
In what follows, $u$ and $d$ quarks are labeled by $i=1-4$, 
and $\bar s$ by $i = 5$.  
The confining potential $V_{\rm Conf}$ is of 
harmonic oscillator type: 
\begin{equation}
V_{\rm Conf} = - \sum_{i<j} \sum_{\alpha=1}^8
 \frac{\lambda^\alpha_i}{2}\,\frac{\lambda^\alpha_j}{2}
 \Big[{k \over 2} \left( {\bf x}_i- {\bf x}_j
\right)^2 + v_0 \Big] \, , 
\label{vconf}
\end{equation}
where ${\bf x}_i$ is a position vector of $i$th quark, $v_0$ 
is a mass shift parameter, and $\lambda^{\alpha}_i$ 
are the Gell-Mann matrices for color; 
note that $\lambda^{\alpha}_i 
\rightarrow -{\tilde \lambda}^{\alpha}_i$ for
antiquark. 
The color-magnetic potential $V_{\rm CM}$ is 
\begin{equation}
V_{\rm CM}=\sum_{i<j}\sum_{\alpha=1}^8
 \frac{\lambda^\alpha_i}{2}\,\frac{\lambda^\alpha_j}{2}
 {{\xi_{\sigma}} \over {m_i m_j}}\,
e^{-( {\bf x}_i-{\bf x}_j )^2/ \beta^2 }
\, \mbox{\boldmath $\sigma$}_i \cdot
  \mbox{\boldmath $\sigma$}_j  \, .
\label{Vcm}
\end{equation}

Parameters in the Hamiltonian are 
fixed as follows. First we take standard values, 
$m_u=m_d=330$ MeV and $m_{\bar s}=500$ MeV, for quark masses. 
And as for $k$ and $\beta$ we take the same values as in 
Refs.~\cite{Isgur71,Isgur83}, i.e. 
$k= 455.1$ MeV$\cdot$ fm$^{-2}$ and $\beta=0.5$ fm. 
The remainder $\xi_{\sigma}$ and $v_0$  
are so determined that 
the three-body calculation, done in the same manner as 
in Ref.~\cite{Hiyama04},
 can reproduce 
$m_N=939$ MeV and $m_\Delta=1232$ MeV. 
The resultant values are 
$\xi_{\sigma}/m_u^2=-474.9$ MeV and $v_0=-428.3$ MeV. 
The parameters thus determined are assumed to be 
universal for all low-lying hadrons 
including the pentaquark.

\begin{table}[htbp]
  \caption{Static properties of conventional mesons and baryons.  
  Squared charge radius in the last columns is defined by 
  $\langle \sum_i Q_i r_i^2 \rangle $ in units of fm$^2$, where 
$Q_i$ and $r_i$ are charge of the $i$th quark 
and distance of the $i$th quark from the center of mass,
respectively. Magnetic moments and squared charge radii 
of $\Delta^Q$ states are 
proportional to the charge $Q$.
  }
  \label{tab:1}
\begin{center}  
  \begin{tabular}{cccccccccc}
\hline
\hline
&  &\multispan2 mass  & &\multispan2 magnetic moment & 
&\multispan2 sq. charge radius \\
&  & Cal.$\;$ &Exp. &  & Cal.  &Exp. & &Cal. &Exp. \\
& &\multispan2 (MeV) & &\multispan2 (nm) &  
&\multispan2 (fm$^2$) \\
\hline 
&$p$     & 939 $\;$  &939 &   &\enskip 2.7735  &\enskip 2.7828  & 
     &\enskip $(0.60)^2$ &\enskip $(0.87)^2$ \\
&$n$    & 939 $\;$  &939  & &$-1.826$  &$-1.913$  & 
     &\enskip$-0.04$ &\enskip$-0.12$ \\
&$\Lambda$   &1058 $\;$  &1115  & &$-0.613$  &$-0.613$   
     & &\enskip $-0.004$ &\enskip -- \\
&$\Sigma^+$  &1119 $\;$  &1189  & &\enskip2.695  &\enskip2.458  
     & &\enskip 0.44 &\enskip -- \\
&$\Sigma^0$  &1119 $\;$ &1192  & &\enskip0.822  &\enskip -- 
     & &\enskip 0.06 &\enskip -- \\
&$\Sigma^-$  &1119 $\;$ &1197  & &$-1.050$  &$-1.160$  
     & &\enskip $-0.31$ &\enskip -- \\
&$\Xi^0$   &1221 $\;$ &1314   & &$-1.466$  &$-1.250$  
     & &\enskip $-0.06$ &\enskip -- \\
&$\Xi^-$  &1221 $\;$ &1321   & &$-0.518$  &$-0.651$  
     & &\enskip $-0.28$ &\enskip -- \\
&$\Delta^Q$ &1232 $\;$  &1232 &  & $2.843Q$  &\enskip --  
     & &\enskip $0.41Q$ &\enskip -- \\
&$\Sigma^{*+}$   &1320 $\;$  &1384 &  &\enskip 3.17  &\enskip -- 
     & &\enskip  0.52   &\enskip -- \\
&$\Sigma^{*0}$   &1320 $\;$ &1384 &  &\enskip 0.32  &\enskip--  
     & &\enskip 0.07 &\enskip -- \\
&$\Sigma^{*-}$   &1320 $\;$  &1384 &  &\enskip $-2.52$  &\enskip-- 
     & &\enskip $-0.38$ &\enskip -- \\
&$\Xi^{*0}$  &1414 $\;$ &1533  & &\enskip 0.64  &\enskip --   
     & &\enskip 0.12 &\enskip -- \\
&$\Xi^{*-}$  &1414 $\;$ &1533  & &\enskip $-2.20$  &\enskip --   
     & &\enskip $-0.35$ &\enskip -- \\
&$\Omega$  &1506 $\;$ &1672 &  &$-1.877$  &$-2.02$ 
 & &\enskip $-0.32$ &\enskip --  \\
\hline 
&$K^{\pm}$  &483  &495 &  &  &  &   & $\pm0.15$  \\
&$K^{0}$ &483  &495 & & & &  & $-0.04$ & \\
&$\pi^{\pm}$  &206  &140 & & & &  & $\pm$0.12& \\
&$\pi^{0}$  &206  &140 & & & &  & 0.0 &\\
&$\rho^{\pm}$ &740  &770 & & & & & $\pm$0.28  &  \\
&$\rho^{0}$ &740  &770 & & & & & 0.0  &  \\
&$\omega$ &740  &783 & &  & & & 0.0 & \\
&$\phi^0$  &939  &1020 &  & & & & 0.0  & \\
&$K^{*\pm}$  &839  &892 &&  & &   & $\pm$0.29  & \\
&$K^{*0}$  &839  &892 & & & &   & $-0.07$  & \\
\hline
\hline
  \end{tabular}
\end{center}  
\end{table}

One may wonder whether  a linear confinement potential 
should be used instead of the harmonic oscillator potential.  
The latter potential, however, has been used in many previous studies, 
since it is easy to handle.  
In our present analysis, rather than the practical advantage, 
we point out that as long as low lying states up to 
$\hbar \omega \sim $ several hundreds MeV are concerned, 
their properties are rather insensitive to the type of the 
potential.  
For this reason, we adopt the harmonic oscillator confining 
potential for the first serious calculation of the 
five-body system.
(Recently, we have applied linear-type
confiment and color-magnetic potential
of ordinary baryons and mesons to
the pentaquark system, and have
confirmed that
our main conclusion of the absence 
of low-lying pentaquark resonances 
is unchanged in this more realistic 
calculation \cite{Hiyama06}.)

The present Hamiltonian is tested 
by static properties of conventional baryons and mesons.  
Results are summarized in Table I which shows theoretical values of 
masses, magnetic moments and charge radii and 
their corresponding experimental 
values. It should be emphasized that 
the present set of interactions satisfactorily reproduces observed
masses of octet and decuplet baryons and octet mesons. 
We also calculated 
absolute strengths of non-leptonic weak decay 
of hyperon and found that 
the results reproduce the observed data 
owing to the $qq$ correlation taken properly
and are very close to those of the previous work
\cite{Hiyama04} in which 
a spin-spin interaction different 
from $V_{\rm CM}$ is taken.  
Thus, agreement of the theoretical values 
with the experimental values for 
the two- and three-quark systems is satisfactory
to proceed to the five-quark system with the 
same Hamiltonian.


\section{Method}

We solve the five-body Schr\"odinger equation
\begin{eqnarray}
( H - E )\, \Psi_{J^\pi M}  = 0 \,,    
\label{schroedinger}
\end{eqnarray} 
imposing the scattering boundary condition to the 
$NK$ scattering component of the total 
wave function $\Psi_{J^\pi M}$; 
here $\Psi_{J^\pi M}$ is classified with 
the angular momentum $J$,
 its $z$-component
$M$ and parity $\pi$.
The $NK$ scattering component of $\Psi_{J^\pi M}$
is expressed by 
\begin{eqnarray}
\hskip -9mm&&\Psi_{J^\pi M} ^{(NK)}(E)  \nonumber \\
\hskip -9mm&&= {\cal A}_{1234}\!
\left\{  \left[ \big[\phi_{\frac{1}{2}}^{(N)}(123)
 \phi_0^{(K)}(45)\big]_{\frac{1}{2}}   
\chi_L({\bf R}_1)\right]_{J^\pi M} \!\right\}, \;
\label{elastic}
\end{eqnarray} 
where the operator ${\cal A}_{1234}$  
antisymmetrizes  particles $1-4$ ($u, d$ quarks)
while particle 5 is ${\bar s}$.
Here, $\phi_{\frac{1}{2}}^{(N)}$ and $\phi_0^{(K)}$
are the color-singlet spin $\frac{1}{2}$ and spin 0 
intrinsic wave functions of $N$ and $K$,
respectively, and   $\chi_L({\bf R}_1)$ is the
wave function of the $NK$ relative motion 
along the coordinate ${\bf R}_1$ 
(cf. $c=1$ of Fig.~\ref{fig:penjacobi})
with the angular momentum $L$
and the center-of-mass energy 
$E- E_{\rm th}$, $E_{\rm th}$ being the 
$NK$ threshold energy (1422 MeV in the present model).
The reported energy of $\Theta^+(1540)$ is slightly 
above the $NK$ threshold, it is then convenient to consider 
the pentaquark energy $E$ with $E- E_{\rm th}$. 

The interaction-region part 
of $\Psi_{J^\pi M}$ should be described with a large model space. 
For this purpose, we take five Jacobi-coordinate set, 
$c=1-5$ of Fig.~\ref{fig:penjacobi}, and construct
the $L^2$-type basis functions, $ \Phi^{(c)}_{J^\pi M,\alpha}$, 
for each coordinate $c$ as follows: 
\begin{eqnarray} 
&&\Phi^{(c)}_{J^\pi M,\alpha}  =
{\cal A}_{1234} \left\{
\xi_{\bf 1}^{(c)}(1234,5) \,
\eta_{0 (t)}^{(c)}(1234,5) \right.
\nonumber \\
&& \!\!\! \times \left.
\left[
\chi_{S (s {\bar s}\sigma)}^{(c)}(1234,5) \:
\psi^{(c)}_{L \{n\}}({\bf r}_c, {\bf \rhovec}_c, {\bf s}_c, {\bf R}_c) 
\right]_{J^\pi M}  \right\}, 
\qquad 
\label{basis} 
\end{eqnarray} 
where  $\xi_{\bf 1}^{(c)}$ is the 
color-singlet wave function, and
$\eta_{0 (t)}^{(c)}$,
$\chi_{S (s {\bar s}\sigma)}^{(c)}$
and
$\psi^{(c)}_{L \{n\}}$  
are  the isospin, spin and 
spatial wave functions 
with the total isospin $T=0$, the total spin $S=\frac{1}{2}$ and the 
total orbital angular momentum $L$, respectively. 
Here we have neglected $S=\frac{3}{2}$, since it is decoupled to 
the $N+K$ scattering channel with $S=\frac{1}{2}$. 
Here, $t$ and $s, {\bar s}, \sigma$ are intermediate 
quantum numbers of isospin and spin coupling, respectively,
and as shown in the next section the symbol $\{n\}$ specifies 
the radial dependence of the spatial wave functions.
The suffix $\alpha$ in $\Phi^{(c)}_{J^\pi M,\alpha}$
specifies a set of
$(t, S, s, {\bar s}, \sigma, L, \{n\})$.

The Hamiltonian is diagonalized within a model space spanned by 
a large number of $\Phi^{(c)}_{J^\pi M,\alpha}$, that is,  
$\sim\!\!15,000$ basis functions in actual calculations. 
The resulting discrete eigenstates are called pseudostates, 
when the eigenenergies $E_\nu$ satisfy $E_\nu > E_{\rm th}$. 
The pseudostates, 
$\{ \widehat{\Phi}_{J^\pi M}(E_\nu) ; \nu=1-\nu_{\rm max} \}$,
are written in terms of  $\Phi^{(c)}_{J^\pi M,\alpha}$ as
\begin{equation} 
\widehat{\Phi}_{J^\pi M}(E_\nu) =\sum_{c, \alpha}
A_{J, \alpha}^{(c)}(E_\nu) \,\,  
\Phi^{(c)}_{J^\pi M,\alpha}({\bf r}_c,
{\bf \rhovec}_c,{\bf s}_c,{\bf R}_c).
\label{expansion}
\end{equation}
It is possible to expand the interaction-region part of the 
total wave function in term of 
those eigenfunctions
since they are considered to form a complete set 
for each $J^\pi$ in the finite interaction region.
The total wave function is then described 
as a superposition of the $NK$ scattering component and 
the $\widehat{\Phi}_{J^\pi M}(E_\nu)$: 
\begin{equation} 
\Psi_{J^\pi M}(E)= \Psi^{(NK)}_{J^\pi M}(E) + 
\sum_{\nu=1}^{\nu_{\rm max}} b_\nu(E)  
\widehat{\Phi}_{J^\pi M}(E_\nu).
\label{fullwave}
\end{equation}
The second term 
describes virtual excitations of
$N$ and $K$ in $c=1$ and other five-body distortions
in $c=2-5$ in the intermediate stage of the scattering.
Unknown quantities 
$\chi_L({\bf R}_1)$ in $\Psi^{(NK)}_{J^\pi M}(E)$
and $ b_\nu(E)$  are obtained by solving 
Eq.~(\ref{schroedinger})
with the Kohn-type variational method~\cite{Kami77}.
Inclusion of high-lying pseudostates in the model space 
do not change calculated values of 
the phase shift, when the energies $E_\nu$ are much larger than 
the $NK$ scattering energy. Hence, 
only a few tens of lowest-lying pseudostates contribute to 
the numerical calculation. 
In the present calculation, the $NK$ scattering energy 
should be smaller than an energy of the 
first spatial excited state of nucleon, 
i.e. the Roper resonance which is located at 
675 MeV above the nucleon mass in the present model, 
because we consider only the $NK$ scattering 
component as an open channel. 
The condition for our calculation to be valid is then 
$E-E_{\rm th} < 675 $ MeV. 
For simplicity, we also ignore 
the $NK^*$ channel, 
since we found that 
inclusion of the channel little 
affects the conclusion of the present work. 
We also analyzed the $J^\pi=\frac{1}{2}^+$ resonance of $S=\frac{3}{2}$
 and $L=1$. 
It is not coupled to the $NK$ scattering channel with $S=\frac{1}{2}$ 
but to the $NK^*$ channel. 
Therefore, it will have a small width, if it appears. 
However, we confirmed that this is not realized at $E-E_{\rm th} < 675 $ MeV. 
These points will be discussed in the forthcoming paper.

\section{Basis functions}

Explicit definitions of the color wave functions 
$\xi_{\bf 1}^{(c)}(1234,5)$ are as follows: 
\begin{eqnarray}
&&\xi_{\bf 1}^{(1)}=\big[(123)_{\bf 1}(45)_{\bf 1} \big]_{\bf 1},\;
\quad \xi_{\bf 1}^{(2)}=\big[[(12)_{\bf {\bar 3}}
(45)_{\bf 1}]_{\bf {\bar 3}}3 \, \big]_{\bf 1},\nonumber\\
&&\xi_{\bf 1}^{(3)}=\big[(12)_{\bf {\bar 3}} 
[(45)_{\bf 1}3]_{\bf 3}\big]_{\bf 1}, \;
\xi_{\bf 1}^{(4)}=\big[[(12)_{\bf {\bar 3}}
(34)_{\bf {\bar 3}}]_{\bf 3}5 \,\big]_{\bf 1}, \nonumber\\
&&\xi_{\bf 1}^{(5)}=\big[(12)_{\bf {\bar 3}}
[(34)_{\bf {\bar 3}}5]_{\bf 3}\big]_{\bf 1}
\end{eqnarray} 
with obvious notation; note that 
$\xi_{\bf 1}^{(1)}=\xi_{\bf 1}^{(2)}=\xi_{\bf 1}^{(3)}$ 
and $\xi_{\bf 1}^{(4)}=\xi_{\bf 1}^{(5)}$ 
due to the recombination of colors. 
The $\xi_{\bf 1}^{(c)}$ are connected for $c=4,5$ and 
molecular for other $c$, 
since there is no color-singlet cluster in the case of 
$c=4,5$.

The isospin wave functions
$\eta_{0 (t)}^{(c)}(1234,5)$ 
with  the total isospin 0 and 
the intermediate isospin $t (=0,1)$ are described as
\begin{eqnarray}
&&\eta_{0 (t)}^{(1)}=
\big[[(12)_t 3]_{\frac{1}{2}}(45)_{\frac{1}{2}}\big]_0, 
\;
\eta_{0 (t)}^{(2)}=
\big[ [(12)_t (45)_{\frac{1}{2}}]_{\frac{1}{2}} 3 \big]_0, 
\nonumber \\
&&\eta_{0 (t)}^{(3)}=
\big[(12)_t [(45)_{\frac{1}{2}} 3]_t\big]_0, \;
\eta_{0 (t)}^{(4)}=\big[[(12)_t (34)_t]_0 5\big]_0,
\nonumber \\
&&\eta_{0 (t)}^{(5)}=\big[(12)_t [(34)_t 5]_t\big]_0. 
\end{eqnarray}
The spin functions 
$\chi_{\frac{1}{2} (s {\bar s}\sigma)}^{(c)}(1234,5)$ 
are also described as
\begin{eqnarray}
&&\!\!\!\!\!\! \chi_{{\frac{1}{2}}(s{\bar s}\sigma)}^{(1)}
\!=\!\big[ [(12)_s 3]_{\sigma}(45)_{\bar s}
\big]_{\frac{1}{2}}, \:
\chi_{{\frac{1}{2}} (s{\bar s}\sigma)}^{(2)}\!=\! 
\big[ [(12)_s (45)_{\bar s}]_\sigma 3 \,\big]_{\frac{1}{2}}
\nonumber \\
&&\!\!\!\!\!\! \chi_{{\frac{1}{2}}(s{\bar s}\sigma)}^{(3)}
\!=\!\big[(12)_s 
[(45)_{\bar s}3\,]_{\sigma} \big]_{\frac{1}{2}}, \: 
\chi_{{\frac{1}{2}} (s{\bar s}\sigma)}^{(4)}\!=\!
\big[ [(12)_s (34)_{\bar s}]_{\sigma} 5\, \big]_{\frac{1}{2}},
\nonumber \\
&&\!\!\!\!\!\! \chi_{{\frac{1}{2}}(s{\bar s} \sigma)}^{(5)}
\!=\!\big[ [(12)_s [(34)_{\bar s} 5]_\sigma \big]_{\frac{1}{2}}.
\end{eqnarray}

Finally, 
the spatial wave function 
$\psi^{(c)}_{L\{n\}}({\bf r}_c, \rhovec_c, {\bf s}_c, {\bf R}_c)$  
with the total orbital angular momentum $L$, where 
$L =0$ for $J^\pi = \frac{1}{2}^-$ and $ L=1 $ 
for $J^\pi= \frac{1}{2}^+$, 
is assumed  as
\begin{eqnarray}
&&\!\!\!\!\!\!\!\!\!\!\!\! 
\psi^{(c)}_{L_c \{n\}}  ({\bf r}_c, 
{\bf \rhovec}_c,{\bf s}_c, {\bf R}_c)  
\nonumber \\ 
&&\!\!\!\!\!\!\!\!\!\!\!\! =  
\phi^{(c)}_{n_r 00}({\bf r}_c)\,  
\phi^{(c)}_{n_\rho 00}(\rhovec_c)\,  
\phi^{(c)}_{n_S 00}({\bf S}_c)\,  
\phi^{(c)}_{n_R L_c M}({\bf R}_c) . 
\label{spatial}
\end{eqnarray}
Here we have set the orbital angular momenta $\{L_{x_c}\}$ associated 
with coordinates ${x_c}={\bf r}_c, {\bf \rhovec}_c, {\bf s}_c$ 
to be zero and the angular momentum $L_c$ associated 
with ${\bf R}_c$ to be $L$; as shown below, however,
this does not mean that the angular momentum space 
is small. The set $ \{ n \}= \{ n_r, n_\rho, n_S , n_R \} $ 
specifies the radial dependence  
of the four basis functions $\phi^{(c)}$. 
In GEM, the functions 
$\phi_{n_R L_c M}({\bf R})$ are written as
\begin{eqnarray}
 \quad \phi_{n_R L_c M}({\bf R}) 
= R^{L_c} e^{-(R/{\bar R}_{n_R})^2}\:
Y_{L_c M }({\widehat {\bf R}})\,  
\label{phibasis}
\end{eqnarray}
with the Gaussian ranges taken in geometric progression, 
\begin{eqnarray}
{\bar R}_{n_R}={\bar R}_1 a^{n_R-1} 
\qquad (n_R=1-n_R^{\rm max})
\label{defRn}
\end{eqnarray}
with $a=({\bar R}_{n_R^{\rm max}}
/{\bar R}_1)^{1/(n_R^{\rm max}-1)}$. 
Here, $n_R^{\rm max}, {\bar R}_1$ and ${\bar R}_{n_R^{\rm max}}$ 
depend upon $c$, isospin and spin taken, but the explicit 
dependence is suppressed for simplicity of notation.
The same procedure is taken also for 
$\phi^{(c)}_{n_r 00}({\bf r}_c)\,,  
\phi^{(c)}_{n_\rho 00}(\rhovec_c)\,$ and  
$\phi^{(c)}_{n_S 00}({\bf S}_c)\,$. 

\vskip 0.1cm
In GEM, 
the model space is constructed by superposing the Gaussian basis functions 
(\ref{spatial}) over 
all $c$ from 1 to 5. This superposition is inevitable to describe 
few-body wave functions accurately, 
particularly when the wave functions have properties of 
strong short-range correlations and/or long-range tails; 
many examples are shown in \cite{Hiyama03}.
As a consequence of the superposition, furthermore, 
the fact that $\{L_{x_c}\}=0$ in (\ref{spatial}) does not mean 
that the angular momentum space is small. As a simple example, let us consider 
a base with $\{L_{x_c}, L_c\}=0$, 
$ 
 e^{-(r_c/{\bar r})^2 - (\rho_c/{\bar \rho})^2
    -(S_c/{\bar S})^2 - (R_c/{\bar R})^2} \;.
$
The base can be rewritten in terms of $c'$, i.e. 
${\bf r}_{c'},\rhovec_{c'},{\bf S}_{c'}$ ${\bf R}_{c'} $. 
The transformed form contains many terms with 
$\{L_{x_c'}, L_{c'}\} \neq 0$. 
Thus, the five-body eigenfunctions
$\widehat{\Phi}_{J^\pi M}(E_\nu) $
given by the superposition can  
cover an angular momentum space large enough to 
derive the phase shifts accurately.

\vskip 0.1cm
The antisymmetrization between quarks $1-4$ 
requires $s+t=$ even for $c=1-3$ and
$s+t={\bar s}+t=$ even for $c=4, 5$.
For $c=1-3$, ${\bar s}=0$ is taken to omit the
$K^*$ component. 
The diquark model proposed for $J^\pi = \frac{1}{2}^+$
in Ref.~\cite{Jaffe03}
is included in our model space as a configuration of $c=4$ with 
$t=s={\bar s}=\sigma=0$ and $L=1$, though the diquark is 
treated as a boson in the model but not in the present approach. 

The calculated phase shifts are converged with 
$n_r^{\rm max}, n_\rho^{\rm max}, n_S^{\rm max}  = 5$ or  6 
for $c=1-5$ and  $n_R^{\rm max}=12$ for $c=1$ and 6 for other $c$. 
Minimum and maximum ranges  of the bases are, respectively,  
0.2 fm and 4.0 fm for coordinate $R_1$ and 
$\sim\!\!~0.2$ fm and $\sim\!\!~2.0$ fm for other coordinates. 
Eventually, the model space is constructed 
by $\sim\!\!~15,000$ basis functions. 

It is worth noting that the connected configuration of $c=4,5$
has a non-negligible overlap with the molecular
configuration of $c=1-3$ when they are localized in a small
space, and in general the overlap is enhanced by the antisymmetrization
${\cal A}_{1234}$.
An extreme case is the 
$(0s)^5$ $J^\pi=\frac{1}{2}^-$ shell-model configuration 
in which the center-of-mass motion is excluded. 
The configuration is equivalent to not only 
a molecular configuration 
\begin{eqnarray}
{\cal A}_{1234}\{ 
\xi^{(1)}_{\bf 1}\, \eta^{(1)}_{0(0)}\,
\chi^{(1)}_{\frac{1}{2}(00 \frac{1}{2})}
e^{-\frac{r_1^2}{4b^2}-\frac{\rho_1^2}{3b^2}
-\frac{3R_1^2}{5b^2}-\frac{s_1^2}{4b^2}} \}
\label{exotic0s5}
\end{eqnarray}
but also to an connected configuration 
\begin{eqnarray}
{\cal A}_{1234}\{ 
\xi^{(4)}_{\bf 1}\, \eta^{(4)}_{0(1)}\,
\chi^{(4)}_{\frac{1}{2}(111)}
e^{-\frac{r_4^2}{4b^2}-\frac{\rho_4^2}{4b^2}
-\frac{R_4^2}{2b^2}-\frac{2s_4^2}{5b^2}}\} \;,
\label{nonexotic0s5}
\end{eqnarray}
where $b=\sqrt{\hbar/m_u\omega}=0.63$ fm and 
$\hbar \omega=(5 \hbar^2 k/3m_u)^{\frac{1}{2}}=300$ MeV.
Thus, the connected and the molecular configuration has 100\% overlap. 
The $(0s)^5$ wave function  is obtained as 
an eigenstate of the approximate Hamiltonian
in which $m_s=m_u$ and $V_{\rm CM}=0$ are taken and 
the color factor 
$ \sum_\alpha \lambda^\alpha_i\,\lambda^\alpha_j$ is replaced 
by its average value $-4/3$.
The expectation value of the full Hamiltonian by 
the $(0s)^5$ $J^\pi=\frac{1}{2}^-$  configuration is  
503 MeV above the $NK$ threshold. 
The $(0s)^5$ configuration is not an eigenstate
of the full Hamiltonian and has a large overlap 
with the $NK$ scattering configuration, so 
it is changed into non-resonant continuum states, 
as shown later, 
when the Schr\"{o}dinger equation is
solved with the $NK$ scattering component.

\section{Results}

Firstly, we calculate the $NK$ elastic 
scattering phase shifts for $J^\pi=\frac{1}{2}^-$ and 
$\frac{1}{2}^+$, by omitting the pseudostate terms in (\ref{fullwave}). 
In this calculation, $N$ and $K$ behave as inert particles.
The resulting phase shifts are
shown in Fig.~\ref{fig:phaseshifts} as dash-dotted lines.
There appears no resonance in the entire energy region.

Next, we do the full-fledged calculation including  the
pseudostate terms in  (\ref{fullwave}).
Calculated phase shifts are shown 
as solid lines.
Up to the energy $E \sim$ 300 MeV, the calculated phase shifts of 
$J^\pi = 1/2^-$ agree qualitatively well with the experimental 
data~\cite{hashimoto,hyslop} 
as well as with the previous quark model 
estimations~\cite{bender}.

No resonance is seen in the energy region $0 - 500$ MeV 
above the $NK$ threshold ($1.4-1.9 $ GeV in mass), 
that is, in the reported energy region of $\Theta^+(1540)$. 
One does see two resonances around 530 MeV; 
one is a sharp $J^\pi=\frac{1}{2}^-$ 
resonance with a width of $\Gamma=0.12$ MeV 
located at $E-E_{\rm th}=540$ MeV
and the other is a broad $\frac{1}{2}^+$ 
resonance at $\sim\!\!~520$ MeV with $\Gamma \sim\!\!~110$ MeV. 
It should be noted here that
our result on the absence of the low-lying
pentaquark resonance is consistent with
some recent lattice QCD results in Refs.
\cite{Mathur04,Takahashi05,Ishii05}, in which, 
in fact, no pentaquark resonance
is observed below about 300 MeV with respect to 
the $N+K$ threshold.
 

%
\begin{figure}[htb]
\begin{center}
\epsfig{file=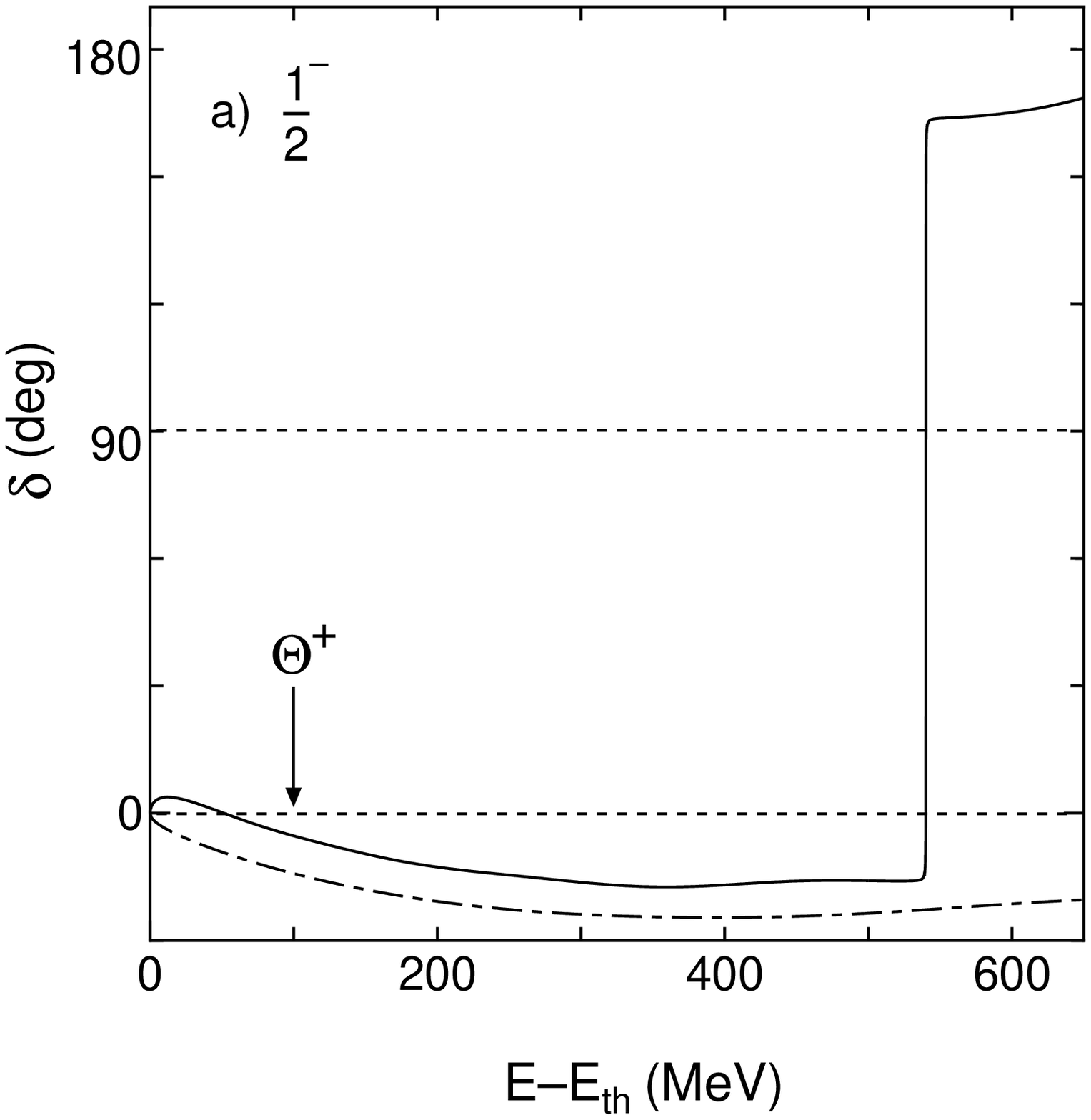,scale=0.40}
\epsfig{file=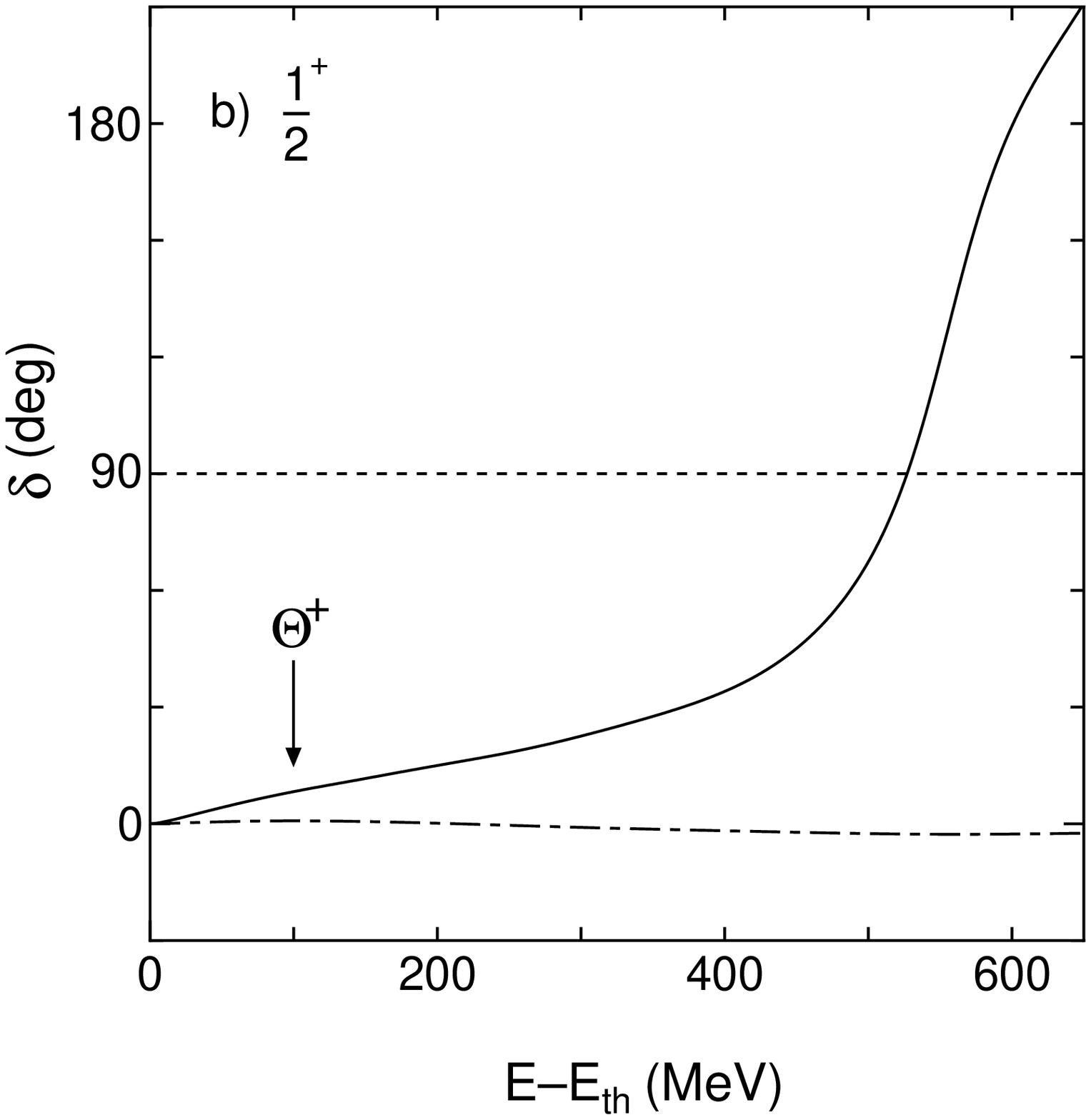,scale=0.40}
\end{center}
\caption{Calculated phase shifts for a) $J^\pi=\frac{1}{2}^-$ and
b) $J^\pi=\frac{1}{2}^+$ states.
The solid lines  are given 
by the full-fledged calculation, while the dash-dotted lines
are by the approximate calculation with the 
elastic $NK$ channel alone (see (\ref{fullwave})).
Energies are measured from the
$NK$ threshold.
The arrow indicates the energy of $\Theta^+(1540)$ 
in $E-E_{\rm th}$.
}
\label{fig:phaseshifts}
\end{figure}
%
\begin{figure}[htb]
\begin{center}
\epsfig{file=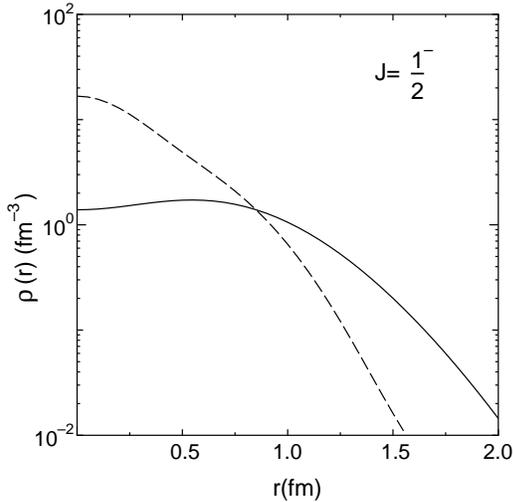,scale=0.4}
\end{center}
\caption{
One-body density $\rho(r)$ of the $J^\pi=\frac{1}{2}^-$ resonance
at $E-E_{\rm th}=540$ MeV as a function of 
$r$ that denotes the distance of a quark
with respect to the center of mass.
The density of $u,d$ quark is given by
the solid line while that of ${\bar s}$ quark
is by the dashed line.
$\rho(r)$ is normalized as $\int_0^\infty
 \rho(r) r^2 {\rm d}r=1$.
}
\label{fig:density}
\end{figure}

We investigate properties of the sharp $J^\pi=\frac{1}{2}^-$ resonance 
by using Eq. (\ref{fullwave}) in its second term. 
In the interaction region which we are interested in,  
the amplitude of the second (pseudostate) term is much larger 
than that of the first (scattering-channel) one. 
Thus, the sharp resonance is not a molecular state of $N$ and $K$ 
in their ground states. 
Figure \ref{fig:density} 
shows the one-body density of the resonance state in its $u, d$ quark  
component and $\bar s$ quark one. One can see from the density that 
a $\bar s$ quark is localized near the center and surrounded by 
$u$, $d$ quarks which are totally antisymmetrized but spatially symmetrized. 
The density of $u, d$ quarks is found to be very broad compared 
with that of the $(0s)^5$ state. This indicates that $u, d$ quarks are not 
in $(0s)$ shell but in higher shells. 
Actually, the overlap of the second term of Eq. (\ref{fullwave}) 
with the $(0s)^5$ configuration is only 2 \%. 
Furthermore, the one-body r.m.s. radius measured from the
center of mass is 1.10 fm for $u, d$ quarks
and 0.72 fm for ${\bar s}$ quark, while the corresponding radius is 
0.69 fm for the $(0s)^5$ configuration.  
Thus, the structure of the sharp resonance is quite far from 
the $(0s)^5$ configuration, although 
the expectation energy of the $(0s)^5$ configuration is
rather close to the resonance energy. 
As for the broad $J^\pi=\frac{1}{2}^+$ resonance, 
the one-body density cannot be estimated meaningfully, 
since the second term are not dominant in Eq. (\ref{fullwave}).

In general neither pure connected nor pure molecular resonance
exists, since any connected and molecular configurations are 
non-orthogonal to each other 
due to the antisymmetrization. 
Nevertheless, we can determine whether the two resonances are 
connected  or molecular in their main components. 
The phase shifts shown above are not changed significantly when 
the connected components (c=4, 5) are omitted from 
the basis (\ref{basis}); for example, the resonance energy 
goes up only by 6 MeV for the $J^\pi=\frac{1}{2}^-$ state and
by 15 MeV for the $J^\pi=\frac{1}{2}^+$ state.
In contrast, when the molecular components $(c=1-3)$ 
are omitted, the $J^\pi=\frac{1}{2}^-$ resonance disappears
from the entire energy region 
and the $J^\pi=\frac{1}{2}^+$ resonance 
is shifted upward by 130 MeV.
This result means that the two resonances are molecular 
in their dominant components. 
Thus, as for both the resonances the five-body configurations 
in the interaction region, represented by the second term of (\ref{fullwave}), 
are mainly molecular, that is, 
they are composed of color-singlet  but spatially distorted (excited)
$(qqq)_{\bf 1}$ and $(q{\bar q})_{\bf 1}$ clusters. 
The reason why the five-body system is 
mainly molecular at energies concerned in this work is as follows. 
Since the $(q{\bar q})_{\bf 1}$ correlation is 
twice as attractive as the $(qq)_{\bf {\bar 3}}$ one, 
in general the molecular configuration 
with one $qq$ pair and one $q{\bar q}$ pair 
(cf. $c=1-3$ in Fig.~\ref{fig:penjacobi})
obtain an energy gain larger than the connected configuration 
with two $qq$ pairs (cf. $c=4, 5$) does. 
Thus, the molecular five-body configuration in which 
the $qq$ and  $q{\bar q}$ correlations
work most effectively appears as low-lying 
states of the pentaquark system. 
For the two resonance states   
$(qqq)_{\bf 1}$ and $(q{\bar q})_{\bf 1}$ clusters are
spatially distorted to a large extent, while for non-resonance states 
the distortion is rather weak. 
Exotic resonances might appear at energies much higher than the two 
molecular resonances do, but it is out of scope of this
paper since some inelastic scattering channels are opened there.

\begin{figure}[htb]
\begin{center}
\epsfig{file=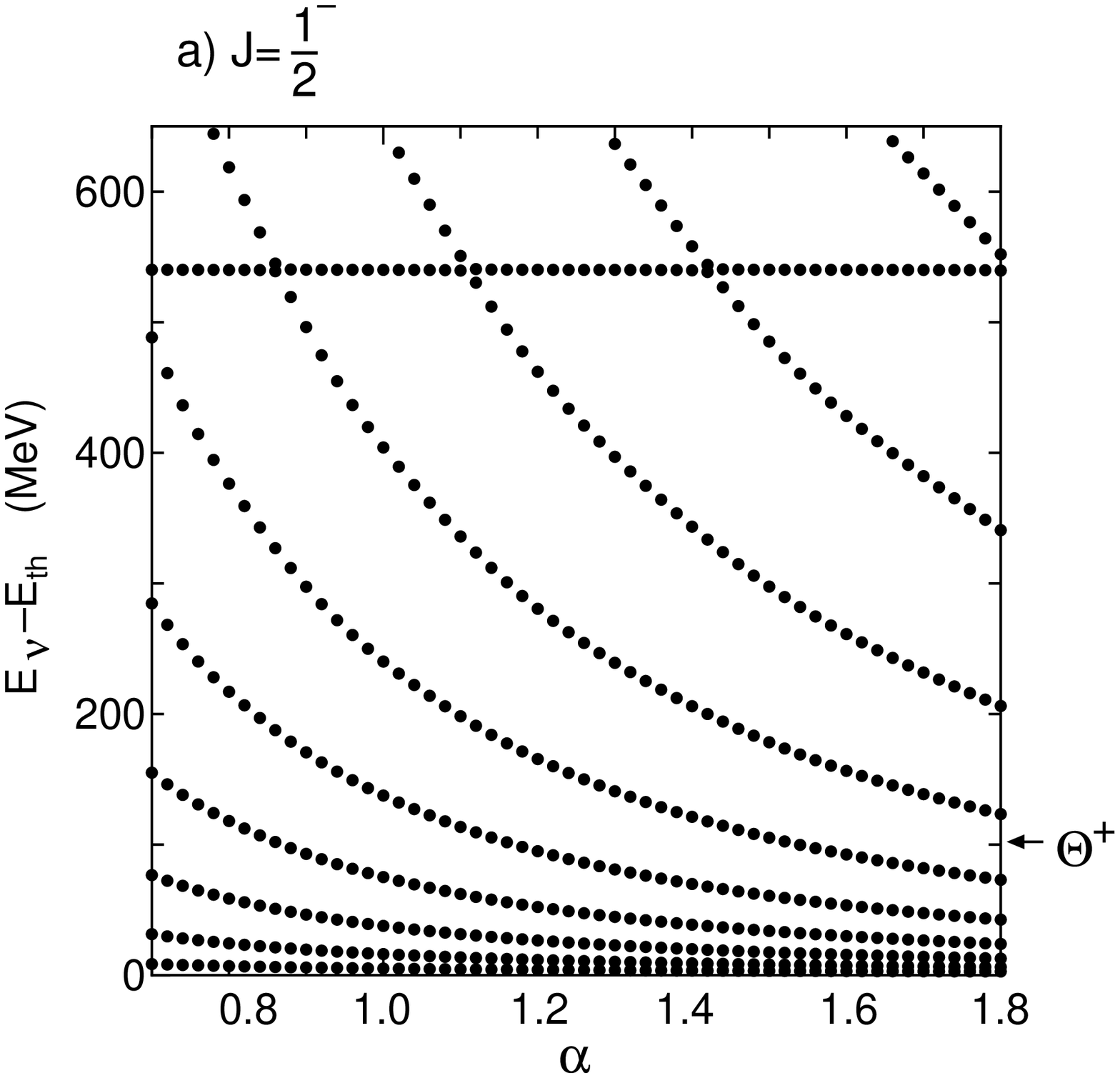,scale=0.38}
\epsfig{file=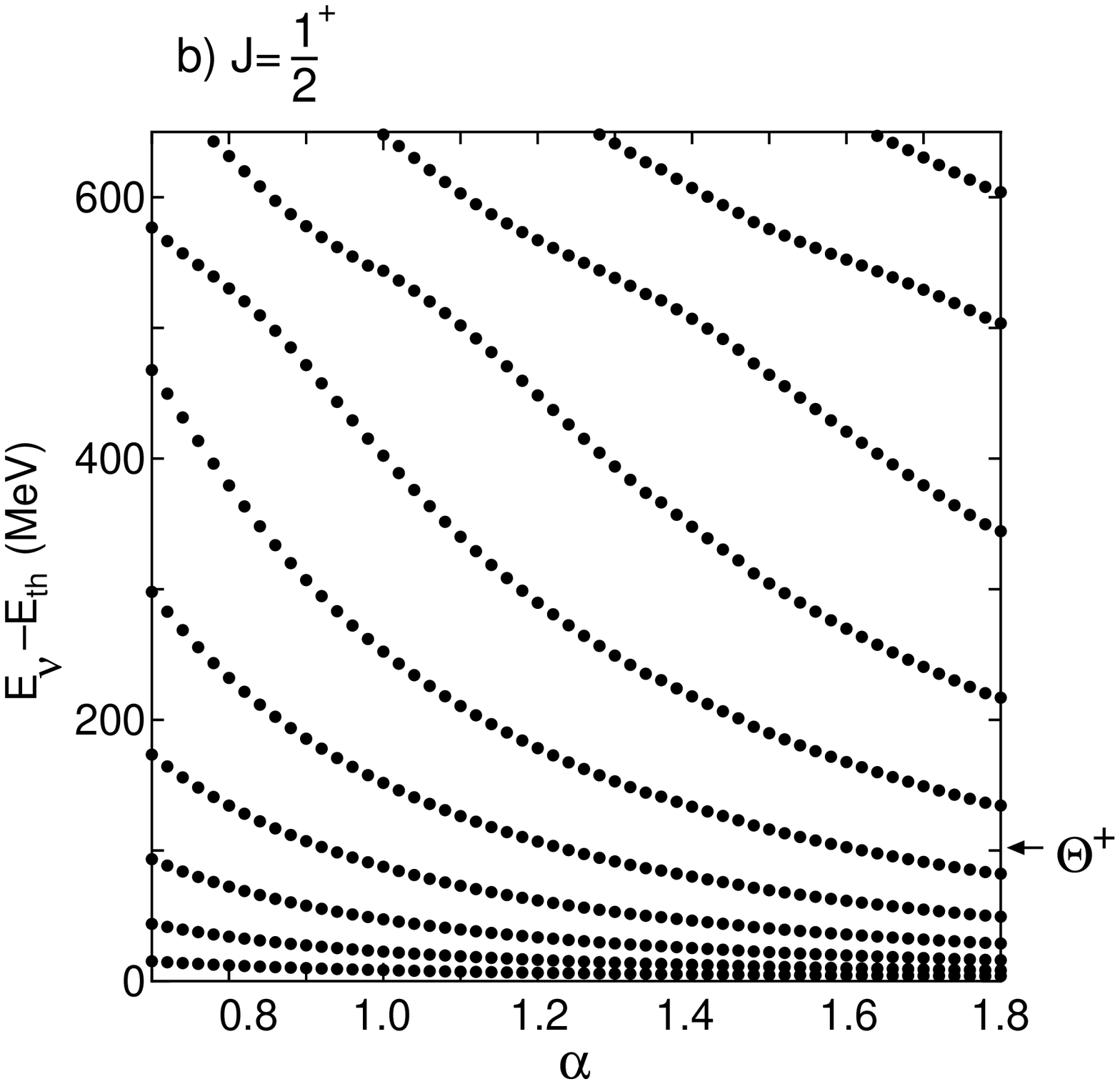,scale=0.38}
\end{center}
\caption{The stabilization plots of the
eigenenergies $E_\nu$ of the pseudostates
against the scaling factor $\alpha$ for 
a) $J^\pi=\frac{1}{2}^-$ and b) $J^\pi=\frac{1}{2}^+$.
Here, the Gaussian ranges ${\bar R}_{n_R}$
associated with the coordinate ${\bf R}_1$
is scaled  as ${\bar R}_{n_R} \rightarrow 
\alpha {\bar R}_{n_R}$ with $0.7 \leq \alpha \leq 1.8$; 
The arrow indicates the energy of $\Theta^+(1540)$
in $E_\nu -E_{\rm th}$.
}
\label{fig:realscaling}
\end{figure}

All the nonrelativistic quark-model calculations done so far 
for the pentaquark system 
use the bound-state approximation in which 
the $NK$ scattering component does not exist asymptotically. 
Therefore, it is of interest
to compare the scattering phase shifts
with the eigenenergies $(E_\nu)$ of the pseudostates as a 
result of the bound-state approximation.
In general, in the case of a sharp resonance, 
there is an eigenenergy $E_\nu$ of the Hamiltonian 
that is close to the resonance energy, 
and the energy $E_\nu$ is stable with respect to
scaling the ranges of the basis functions 
associated with ${\bf R}_1$ the coordinate of 
the $NK$ scattering, 
since  the asymptotic amplitude of the sharp resonance
is relatively much smaller than the internal amplitude.
In contrast, other pseudostates have eigenenergies $E_\nu$ 
which decrease monotonously with the increasing the ranges, because 
so does the kinetic energy associated with ${\bf R}_1$. 

These characteristics are really seen 
in Fig.~\ref{fig:realscaling} that shows
stabilization plots of $E_\nu$ 
with respect to the scaling $R_{n_R} \rightarrow \alpha R_{n_R}$; 
the parameter $\alpha$ corresponds to the volume size in the lattice 
calculation. 
This plotting to investigate resonances 
is called the real-scaling (stabilization) method \cite{Simons81}. 
In Fig.~\ref{fig:realscaling}a for $J^\pi=\frac{1}{2}^-$, 
all pseudostates except one are unstable in the sense that
they decrease toward the
$NK$ threshold. Thus, those pseudostates are regarded as 
a discrete representation of the non-resonant continuum spectrum.
The stable (horizontal) line at $E_\nu=$ 540 MeV
precisely corresponds to the sharp resonance at 540 MeV
in Fig.~\ref{fig:phaseshifts}a. 
The horizontal line and the nearby unstable lines 
do not cross each other because of repulsive forces 
working between them, though it is not precisely plotted. 
This is called the avoiding crossing. 
One can roughly estimate the width of the 
resonance from the behavior of the avoiding crossing 
with the aid of Eq.~(4) of Ref.~\cite{Simons81}. 
In the present case, the estimated width is of 
order 0.1 MeV and consistent with the value obtained from the 
phase shift.
On the other hand, the $J^\pi=\frac{1}{2}^+$ 
resonance at $\sim\!\!~520$ MeV 
with $\Gamma \sim\!\!~110$ MeV
in Fig.~\ref{fig:phaseshifts}b
is too broad to be identified as a resonance
from the stabilization plot of Fig.~\ref{fig:realscaling}b, 
though one sees a tendency of plateaus around $500-600$ MeV. 
Therefore, without the scattering calculation
it is difficult to  discriminate a broad resonance 
from non-resonance states.

\vskip 0.2cm

In summary, we solved the five-quark scattering problem  
by applying GEM \cite{Hiyama03} 
and the Kohn-type variational method \cite{Kami77} 
to the large model space 
including the $NK$ scattering component. 
We adopted the standard non-relativistic quark model 
of Isgur-Karl which satisfactorily reproduces experimental values 
of the two- and three-quark systems. 
The resultant $NK$ scattering phase shift showed no resonance 
in the reported energy region of $\Theta^+(1540)$;
this is the most important result of the present paper. 
At energies much higher (by $\sim\!\!~400$ MeV)
than the $\Theta^+(1540)$ energy, 
we did find a broad $J^\pi=\frac{1}{2}^+$ resonance 
with $\Gamma \sim\!\!~110$ MeV 
at $\sim\!\!~520$ MeV above the $NK$ threshold
and a sharp $J^\pi=\frac{1}{2}^-$ resonance with $\Gamma=$
0.12 MeV at 540 MeV.  
In the present model Hamiltonian, since the 
$J^\pi=\frac{1}{2}^+$ and $\frac{3}{2}^+$ states are degenerate to 
each other,  there exists a broad $J^\pi=\frac{3}{2}^+$ resonance with 
the same energy and width as the $J^\pi=\frac{1}{2}^+$ resonance. 

We have done the same calculation 
for other Hamiltonian 
proposed in \cite{Hiyama04} that 
also reproduces the observed properties of
the ordinary hadrons and mesons with the same quality of
agreement as in Section II. 
The result was qualitatively the same as in this paper; 
in particular resonances are absent in the low energy 
region up to 500 MeV from the $NK$ threshold.   
The locations and the widths of the resonances at higher energies, 
however, depend on the details of the model hamiltonian.  

The resonance states are mainly composed of color-singlet 
$(qqq)_{\bf 1}$ and $(q{\bar q})_{\bf 1}$ clusters which 
are distorted (excited) to a large extent. 
Thus, these are molecular resonances.  
The $(q{\bar q})_{\bf 1}$ correlation is twice as attractive as 
the $(qq)_{\bf {\bar 3}}$ one, so low-lying states, 
no matter whether the widths are small or not, are dominated by 
molecular configurations. Thus, even if a resonance 
with a small width is measured, the fact that 
the width is small does not necessarily mean that the resonance is connected.
The sharp $J^\pi=\frac{1}{2}^-$ resonance has a quite different structure from 
the $(0s)^5$ configuration. In the resonance, $\bar s$ quark is located near 
the center and surrounded by $u, d$ quarks which are spatially symmetrized.

Finally, we tested 
the reliability of the bound-state approximation with
the real scaling method. 
Essentially the same approximation 
is used in the lattice calculation. 
We found, by comparing Fig. 4 with Fig. 2, that  
the approximation surely works for a sharp resonance with a width 
of order 0.1 MeV, but not for a broad resonance with a width 
of order 100 MeV and that most of the states obtained by the
approximation melt into non-resonant continuum states
when the $NK$ scattering channel is included accurately. 
It should be noted that our model calculation with the $NK$
scattering channel clarifies the mechanism of the disappearance
of low-lying pentaquark resonace, 
which cannot be shown even by lattice QCD.
Further analyses including $NK^*$ scattering channel 
will be reported in a forthcoming paper, 
together with results for other $(T, J^\pi)$ states.

\vskip 0.3cm

The authors would like to thank Prof. M. Oka, 
Prof. H. Suganuma, Prof. M. Ohbu
and Prof. M. Tanifuji
for helpful discussions. This work has been supported
in part by the Grant-in-Aid for Scientific Research
(14540271) of Monbukagakushou of Japan and by
the Nara Women's University Intramural Grant for Project
Research (E.H.).
The numerical calculations were done on FUJITSU VPP5000 at
JAERI.

\end{document}